\documentclass[doublecol]{epl2}

\usepackage{amsmath}
\usepackage{amssymb}
\usepackage{graphicx}
\usepackage{color}

\title{Satellite probing General Relativity and its extensions and Kolmogorov analysis}
\shorttitle{Satellite probing General Relativity and its extensions and Kolmogorov analysis}

\author{V.G.Gurzadyan$^1$, I.Ciufolini$^{2,3}$, S.Sargsyan$^1$, G.Yegorian$^1$, S.Mirzoyan$^1$, A.Paolozzi$^4$}

\shortauthor{V.G.Gurzadyan \etal}

\institute{1. Center for Cosmology and Astrophysics, Alikhanian National Laboratory and Yerevan State University, Yerevan, Armenia; 2.Dipartimento di Ingegneria dell'Innovazione, University of Salento, Lecce; 3.Centro fermi, Rome, Italy; 4.Scuola di Ingegneria Aerospaziale and DIAEE, Sapienza University, Rome, Italy}


\abstract{We apply the Kolmogorov statistic to analyse the residual data of two LAGEOS satellites on General Relativistic Lense-Thirring effect,
and show that it reveals a tiny difference in the properties of the satellites, possibly related to Yarkovsky-Rubincam effect. The recently launched LAser RElativity Satellite (LARES) can provide constraints to the extensions of General Relativity such as the Chern-Simons (CS) gravity with metric coupled to a scalar field through the Pontryagin density, so an explicit dependence on the frame dragging measurements vs the CS parameter is given.
}

\begin{document}

\maketitle



\section{Introduction}

The satellite testing of General Relativity was instrumental in confirming by now its predictions, particularly, on the Lense-Thirring (LT) effect \cite{CP,C}. Even higher-accuracy probing of LT effect is expected by means of the LARES satellite currently on nearly zero-eccentricity geocentric orbit \cite{Lares,Lares1}. The importance of such high accuracy tests is increased due to indications of the accelerated expansion of the Universe and the puzzle of the dark energy, for which various models have been proposed including extensions of General Relativity. Chern-Simons (CS) gravity which follows from the string theory, is among the discussed ones \cite{Smith,YYT}, and LARES can improve the constraints on its parameters.

In the present note we apply the Kolmogorov's statistic \cite{Kolm,Arnold,Arnold_MMS} for the first time to satellite LT data, i.e. to the residual data of the two LAser GEOdynamics Satellites (LAGEOS); also, this illustrates the method, before the LARES data are available. This method which enables studying the correlations vs the degree of randomness in a sequence of numbers, has been already applied for the analysis of the cosmic microwave background (CMB) data of Wilkinson Microwave Anisotropy Probe (WMAP) \cite{GK_KSP,K_sky}. That approach has revealed, among other issues, the enhanced degree of randomness of the Cold Spot, a non-Gaussian region in the CMB sky, thus supporting its void nature in the large scale matter distribution; for recent discussion of the Cold Spot by the Planck Collaboration see \cite{Planck}.
  
Then, we inquire into the explicit quantitative values of the Chern-Simons parameter upon the expected increase of accuracy of measurements by LARES.

\section{Kolmogorov analysis of LAGEOS and LAGEOS 2 residuals}

Before turning to the analysis of the data of the two satellites, we briefly introduce the Kolmogorov statistic \cite{Kolm,Arnold,Arnold_MMS}. 
The definition of the Kolmogorov parameter  for a finite random sequence of real numbers $x_1, x_2, \ldots , x_n$ sorted in increasing manner  $x_1 \leq x_2 \leq \ldots \leq x_n$ includes the empirical distribution function 
\begin{equation}
F_n(X)= \left\{
\begin{array}{rl}
	0, & X < x_1 \\
	k / n, & x_k \leq X < x_{k+1} \\
	1, & x_n \leq X\\
\end{array}
\right.
\label{eq:empiricdistribution}
\end{equation}
and the theoretical cumulative distribution function (CDF) 
$$F(X) = n \cdot(probability\, that\, x \leq X).$$

The parameter $\lambda_n$  is defined as
\begin{equation}\label{KSP}
\lambda_n=\sqrt{n}\ \sup_x|F_n(x)-F(x)|\,
\end{equation}
which also is a random variable.

Kolmogorov's theorem states that the probability 
\begin{equation}
\lim_{n\to\infty}P\{\lambda_n\le\lambda\}=\Phi(\lambda)\ ,
\end{equation}
is uniformly converging  at $n \rightarrow \infty$ to $\Phi(\lambda)$: 
\begin{equation}
\Phi(\lambda)=\sum_{k=-\infty}^{+\infty}\ (-1)^k\ e^{-2k^2\lambda^2},\, \Phi(0)=0,\,   \lambda>0\ ,\label{Phi}
\end{equation}
for any continuous cumulative distribution function. $\Phi(\lambda)$ is a monotonic function and varies within $\Phi(0)=0$ to $\Phi(\infty)=1$.

We will use this method to analyze the degree of randomness in the residual data, i.e. the difference between the calculated and the observed positions (in angular degrees) of the two LAGEOS satellites. The satellites have been launched on 4 May 1976 (LAGEOS) and 23 October 1992 (LAGEOS 2) and the data have been collected during nearly 11 years (about 4018 days), with temporal step t=14 days; for details see \cite{CP,C} and refs therein. 

We have computed the function $\Phi$ for the residuals of both LAGEOS satellites' datasets for Gaussian CDF vs the variation of the standard deviation $d\sigma$ of the latter (Figure 1).
We see that there is a difference in the behaviour of $\Phi$ for LAGEOS and LAGEOS 2, i.e. the values for the former do reveal an
enhanced randomness (about 10 times, cf.\cite{Arnold1}) in fitting the Gaussian with respect to those of LAGEOS 2. This indicates that the method is sensitive to a difference in the residual randomness. 

\begin{figure}[h]
\begin{minipage}[v]{0.49\linewidth}
\center{\includegraphics[width=1.0\linewidth]{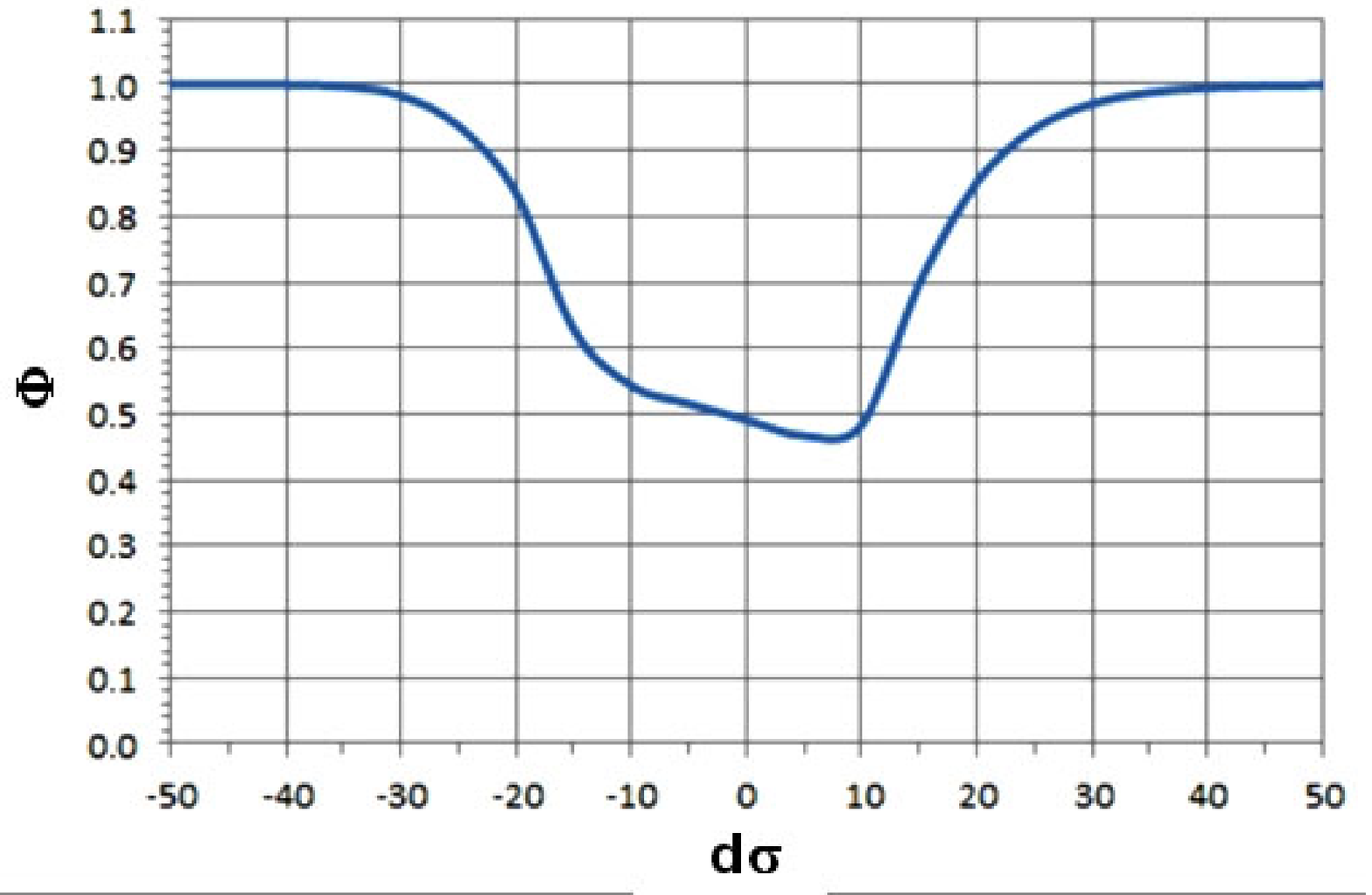} \\ (a)}
\end{minipage}
\hfill
\begin{minipage}[v]{0.49\linewidth}
\center{\includegraphics[width=1.0\linewidth]{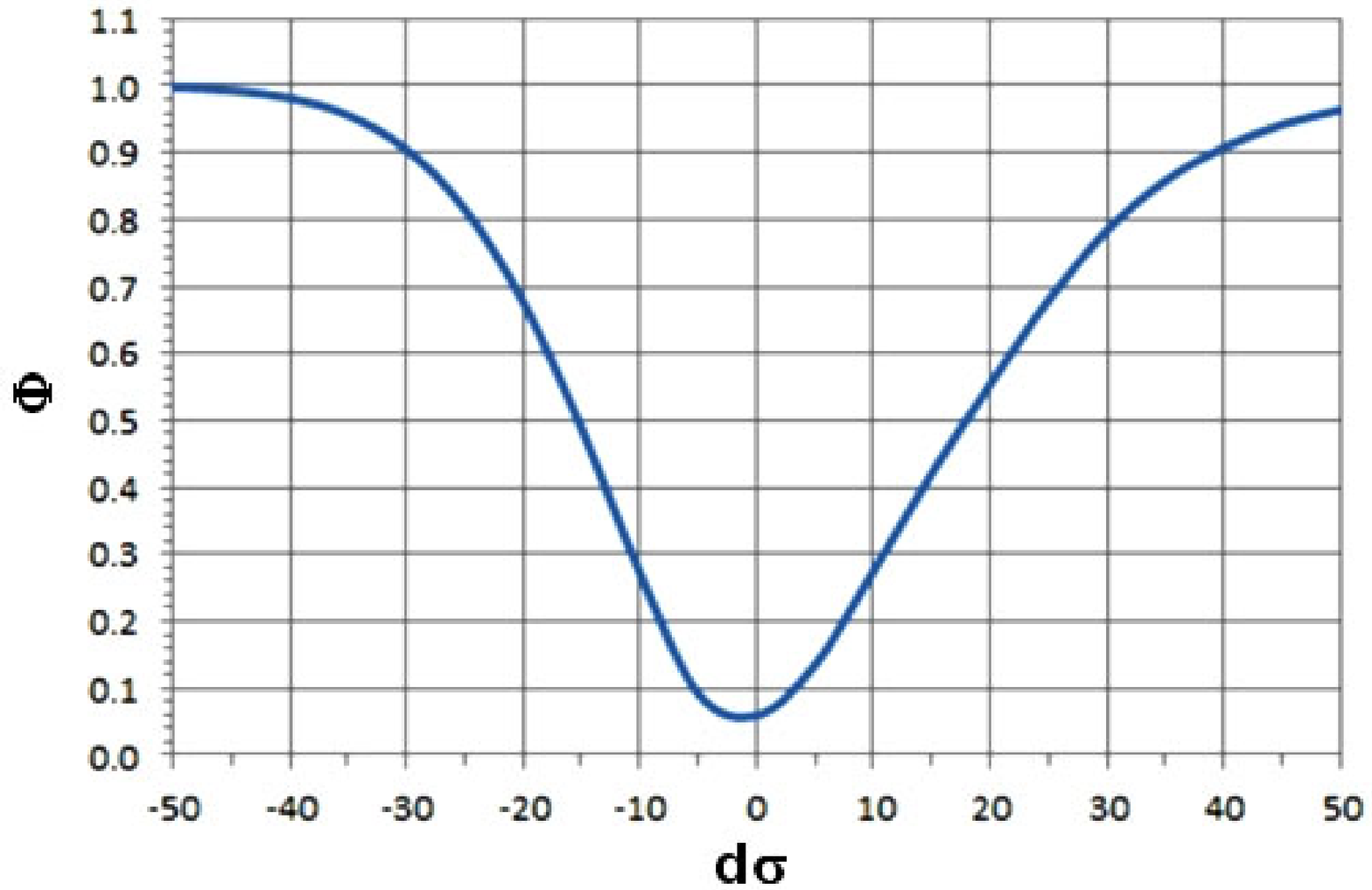} \\ (b)}
\end{minipage}
\caption{Plot the function $\Phi$ vs $d\sigma$ for the residuals of LAGEOS  (a) and LAGEOS 2 (b).}
\end{figure}

Since the LAGEOS and LAGEOS 2 satellites are basically identical, why is there such a difference in the randomness of their orbital residuals?
The explanation of the difference in the randomness of the two sets of residuals might be found in the unmodelled orbital perturbations of the two satellites due to an effect known as thermal thrust or Yarkovsky effect. Let us first explain the origin of that perturbation.
The electromagnetic radiation from the Sun and the radiation from Earth, each instantaneously heat one hemisphere of LAGEOS and LAGEOS 2. Because of the finite heat conductivity of their body, there is an anisotropic distribution of temperature on each satellite. Thus, according to the Stefan-Boltzmann law, there is an anisotropic flux of energy ($\sim \, T^4$) and momentum from the surface of the satellite giving rise to its acceleration. However, if the satellite is spinning fast enough, the anisotropy in the satellite temperature distribution is mainly latitudinal and therefore the acceleration is directed along the spin axis of the satellite. 
The acceleration, \(a_{\Delta \, T}\), due to this `'thermal thrust'', acting along the satellite spin axis, is thus
\begin{equation}
a_{\Delta \, T} \, \sim \, {4 \varepsilon \, \pi r^2_L \, \sigma \, T^3 \, \Delta \, T \over cm_L},
\end{equation} 
  where \(\varepsilon \, \cong \, 0.4\) is the emissivity coefficient of LAGEOS, \(r_L \, = \, 30\) cm its radius, \(m_L \, = \, 4.1 \, \times \, 10^5\) g its mass, \(\sigma = 5.67 \, \times \, 10^{-8}\) W m\(^{-2}\) K\(^{-4}\) Stefan's constant, \(T\) the average equilibrium temperature of LAGEOS and \(\Delta \, T\) the temperature gradient between the two  hemispheres. For example, for \(T \, \cong \, 280\) K and \(\Delta \, T \, = \, 5\) K, we get an acceleration \(a_{\Delta \, T}\) of the order of \(a_{\Delta \, T} \, \sim \,2 \cdot  10^{-9}\) cm/sec\(^{2}\) that, by assuming that the satellite spin axis is substantially constant along one orbit, is constant in direction along one LAGEOS or LAGEOS 2, orbit. To see the effect of this acceleration on the node of LAGEOS or LAGEOS 2, we insert this value of the 'thermal acceleration' in the equation for the rate of change of the nodal longitude
\begin{equation}
{d \, \Omega \over dt} \, = \, {1 \over nasin \, I} \, \left (1 - e^2 \right )^{- {1 \over 2}} \, f \, W \, {r \over a} \, sin \, ( \, \nu \, + \, \omega \, ), 
\end{equation}
where \(n \, = \, 2 \pi / P\) is the orbital mean motion, \(P \, = \) 3.758 h the LAGEOS orbital period, \(a \, = \) 12,271 km its orbital semi-major axis, \(I \, \cong \, 109^{\circ}\).94 its orbital inclination, \(e \, = \, 4 \, \times \, 10^{-3}\) its orbital eccentricity, \(f\) the magnitude of the external force per unit satellite mass, \(W\) the direction cosine of the force \(f\) along the normal to the orbital plane, \(\nu\) the true anomaly and \(\omega\) the argument of the pericenter. 

Finally, assuming \(W\) be constant along one orbit, we can simply integrate this expression over one orbital period, \(P\) = 3.758 h.
Furthermore, there is a variant of the Yarkovsky effect observed on the LAGEOS satellites, called Earth-Yarkovsky or Yarkovsky-Rubincam effect \cite{R}. Rubincam discovered that the infrared radiation from Earth plus the thermal inertia of the LAGEOS retro-reflectors can cause a force on the satellite. Infrared radiation from Earth is absorbed by the LAGEOS retro-reflectors, therefore, due to their thermal inertia and to the rotation of the satellite, there is a LAGEOS, and LAGEOS 2, latitudinal temperature gradient. The corresponding thermal radiation causes an acceleration with an along-track component opposite to the satellite motion. This anisotropic thermal radiation may cause an acceleration of the LAGEOS satellites of the order of 10$^{-10}$ $cm/s^2$.

The orbits of LAGEOS and LAGEOS 2 have approximately the same semi-major axis, that of LAGEOS is equal to 12270 km and that of LAGEOS 2 to 12160 km, thus they approximately have the same orbital period (Kepler's third law). However, the two LAGEOS satellites have a different orbital inclination (the angle between their orbital plane and the Earth equatorial plane), equal to about 110 $^\circ$ for LAGEOS and to about 52$^\circ$  for LAGEOS 2. The frequency of precession of the satellites' orbital plane, relative to 'inertial' space, depends on the satellites' orbital inclination (and is owed to the quadrupole moment and to higher multipole moments of Earth), thus, since the inclinations of the two LAGEOS satellites are different, the rates of change of the orientation of their orbital plane and the frequencies of change of their nodal line (that is, the intersections of their orbital planes with the Earth equatorial plane) are different for the two satellites. Therefore, since the thermal thrust perturbations of their nodal line depend on the relative orientation of satellites' spin axis, satellites' orbital plane and Sun, thermal thrust will perturb the nodal line of each of the LAGEOS satellite with different frequencies.
So why do we observe a difference in the randomness of the residuals of LAGEOS and LAGEOS 2?

A possible answer is that in the previous description of the 'thermal accelerations', both solar Yarkovsky and Earth Yarkovsky, we have assumed a constant spin axis direction of LAGEOS and LAGEOS 2, and a fast enough (with respect to the orbital period) spin rate necessary for the latitudinal thermal gradient to be built between the north and the south hemispheres. However, we carried out analysis of the LAGEOS and LAGEOS 2 orbits and of their residuals between 1992 and 2003. During this period, the conditions of constant spin axis orientation and fast enough spin rate were satisfied by LAGEOS 2 launched in 1992, but not by LAGEOS launched back in 1976. In other words, during the period of our analysis, the LAGEOS satellite had an extremely low spin rate and its spin axis orientation, far from being constant, was almost chaotic (random). 
In conclusion, the LAGEOS satellite should not show any periodical effects in the nodal residuals due to thermal accelerations. On the other hand LAGEOS 2 was spinning fast enough during that period of analysis and with constant spin orientation, so it should show regular periodical effects in the orbital residuals.


\section{Chern-Simons gravity vs LARES's forthcoming measurement accuracy}

Chern-Simons gravity is given by the second-order modification of the Einstein-Hilbert action in the form (see \cite{Smith,YYT}) and refs therein)
\begin{eqnarray}
     S=\frac{1}{16\pi k}\int {\rm d}^4 x \sqrt{-g} \nonumber \times \\
     \left[R+\frac{l}{4} \theta R\overline{R}- 
		 \frac{1}{2} (\partial \theta)^2- V(\theta) +{\cal L}_{\mathrm{mat}}\right],
\end{eqnarray}
where $R$ is the Ricci scalar, $\theta$ is a scalar field, $l$ is a coupling constant. 
$R\overline{R}$ is known as the Pontryagin density, which has the following form
\begin{equation}
R\overline{R}=R_{a}^{bcd}\overline{R}_{bcd}^{a},
\end{equation}
where the dual of the Riemann tensor is
\begin{equation}
\overline{R}_{bcd}^{a}=\frac{1}{2}\epsilon _{efcd}R_{b}^{aef},
\end{equation}
with $\epsilon_{efcd}$, the 4-dimensional Levi-Civita tensor.

The following is denoted as topological Pontryagin current
\begin{equation}
K^{a}=\epsilon ^{abcd}\left( \Gamma _{bk}^{l}\partial _{c}\Gamma _{dl}^{k}+%
\frac{2}{3}\Gamma _{bk}^{l}\Gamma _{cl}^{n}\Gamma _{dn}^{k}\right),
 \label{current}
\end{equation}
satisfying the relation 
\begin{equation}
\Delta _{a}K^{a}=\frac{1}{2}R\overline{R}
\end{equation}
and (\ref{current}) turns to
\begin{eqnarray}
K^{a}=\epsilon ^{abcd}\Gamma _{bk}^{l}\left( \partial _{c}\Gamma _{dl}^{k}+%
\frac{2}{3}\Gamma _{cl}^{n}\Gamma _{dn}^{k}\right) = \nonumber\\
\epsilon ^{abcd}\Gamma
_{bk}^{l}\left( \frac{1}{2}R_{cdl}^{k}-\frac{1}{3}\Gamma _{cl}^{n}\Gamma
_{dn}^{k}\right)
\end{eqnarray}

The variation of the action with respect to the metric gives 
\begin{eqnarray}
     \delta S=\frac{1}{16\pi k}\int d^{4}x\sqrt{-g} 
     \left[ \left( R^{ab}-\frac{1}{2}g^{ab}R+lC^{ab}-\right.\right.\nonumber\\   
\left.8\pi kT^{ab}\biggr)
 \delta g^{ab}+\left( \frac{1}{4}R\overline{R}+g^{ab}\Delta _{a}\Delta _{b}\theta 
-\frac{dV}{d\theta }\right)\delta \theta
+\Sigma \right].
\end{eqnarray}

The equation of motion for the scalar field $\theta$ is
\begin{equation}
     \Box \theta = \frac{{\rm d} V}{{\rm d}\theta}-\frac{1}{4} l R\overline{R} .
\label{scalar}
\end{equation}

Then the modified gravitational field equations are
\begin{equation}
G^{ab}+lC^{ab}=8\pi T^{ab}, 
\label{field_equation}
\end{equation}
where $G^{ab}$ is the Einstein tensor and $C^{ab}$ is the Cotton-York tensor defined as
\begin{equation}
C^{ab}=v_{l}\left( \epsilon^{lacd}\nabla _{c}R_{d}^{b}+\epsilon
^{lbcd}\nabla _{c}R_{d}^{a}\right) +v_{lk}\left( \overline{R}%
^{kalb}+\overline{R}^{kbla}\right).
\end{equation}
and $v_{l}$ and $v_{lk}$ are Chern-Simons velocity and acceleration
\begin{equation}
v_{l}=\partial _{l}\theta =\nabla _{l}\theta, 
\end{equation}
\begin{equation}
v_{lk}=\nabla _{l}v_{k}=\nabla _{l}\nabla _{k}\theta.
\end{equation}
and $T^{ab}$ is combined from the matter stress-energy tensor $T_{mat}^{ab}$
and scalar field stress-energy tensor $T_{\theta }^{ab}$. The latter has the following form
\begin{equation}
T_{\theta }^{ab}=\left( \nabla _{a}\theta \right) \left( \nabla _{b}\theta
\right) -\frac{1}{2}g_{ab}\left( \nabla _{a}\theta \right) \left( \nabla
^{a}\theta \right) -g_{ab}V\left( \theta \right)
\end{equation}


\begin{figure}[h]
\includegraphics[width=20pc]{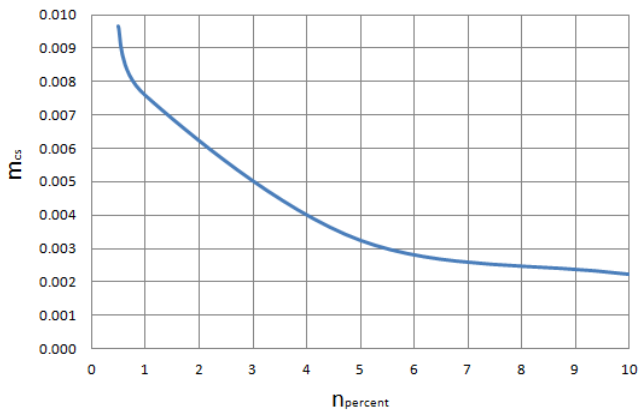}\hspace{2pc}%
\caption{Plot of Chern-Simons parameter $m_{\rm CS}$ (in km$^{-1}$) vs the Lense-Thirring measurement accuracy ratio $n$ (in percents).}
\label{fig:m}
\end{figure}

The Lense–Thirring secular drag rate of the node for a test particle freely orbiting a central rotating mass is \cite{C}
\begin{equation}
     \dot{\Omega}_{\rm GR} = \frac{2 G L}{a^3(1-e^2)^{3/2}},
\end{equation}
where $L$ is the angular momentum of the central mass and $e$ is the orbital eccentricity of the test particle, i.e. the satellite.
The ratio of the drag rates for Chern-Simons theory and for General Relativity is \cite{Smith}
\begin{equation}
     \frac{\dot{\Omega}_{\rm CS}}{ \dot{\Omega}_{\rm GR} } = 15 \frac{a^2}{R^2}
     j_2(m_{\rm CS} R) y_1(m_{\rm CS} a),
\end{equation}
where $j_\ell(x)$ and $y_\ell(x)$ are the first and the second kind spherical Bessel functions, respectively, and $m_{CS}=-3/lk^2{\theta}^2$.

LAGEOS satellites being on practically identical orbits with semi-major axes $a = 12 271$ km verified the General Relativistic Lense-Thirring effect in the gravitational field of the Earth with an accuracy of about 10\% \cite{CP}. Since the LARES is expected to produce higher accuracy data,  
\begin{equation}
\dot{\Omega}_{\rm LARES}= \dot{\Omega}_{\rm GR} (1+n),
\end{equation}
we represent the plot for $n=\frac{\dot{\Omega}_{\rm CS}}{\dot{\Omega}_{\rm GR}}$ (in percents) vs the lower limit of the Chern-Simons parameter $m_{\rm CS}$ (in km$^{-1}$) in Figure \ref{fig:m}.

\section{Conclusions}

Kolmogorov statistic is used for the analysis of the residuals of the satellites LAGEOS and LAGEOS 2, while measuring the Lense-Thirring effect in Earth's gravity. A slight difference in the behaviour of $\Phi$ for the two satellites is revealed, which can be due to non-gravitational effect, i.e. Yarkovsky-Rubincam thermal thrust at differently spinning satellites with non-equal stay time in the orbit. This is the first use of this method for satellite measurements on gravitomagnetism, and later it may be applied to higher-accuracy data by LARES. Chern-Simons gravity parameter's dependence on the increasing accuracy of the measurements expected by LARES is exhibited. The empirical constraints on CS theory are important also due to the attempts to use it to explain the properties of the dark energy, the expansion of the Universe and the processes in galactic nuclei.    

\section{Acknowledgments}

We are thankful to the referee for helpful comments. I. Ciufolini and A. Paolozzi acknowledge the Italian Space Agency for the contract n. I/034/12/0.

\end{document}